-------------- Enclosure number 1 ----------------
\documentstyle[12pt]{article}
\begin{document} 
\global\parskip 6pt
\newcommand{\be}{\begin{equation}}
\newcommand{\ee}{\end{equation}}
\newcommand{\bea}{\begin{eqnarray}}
\newcommand{\eea}{\end{eqnarray}}
\newcommand{\non}{\nonumber}

\begin{titlepage}
\hfill{hep-th/9908150}
\vspace*{1cm}
\begin{center}
{\Large\bf Gott Time Machines, BTZ Black Hole Formation,}\\
\vspace*{.2cm}
{\Large \bf and Choptuik Scaling}\\
\vspace*{2cm}
Danny Birmingham\footnote{E-mail: dannyb@pop3.ucd.ie}\\
\vspace*{.5cm}
{\em Department of Mathematical Physics\\
University College Dublin\\
Belfield, Dublin 4, Ireland}\\
\vspace*{1cm}
Siddhartha Sen\footnote{E-mail:
sen@maths.tcd.ie}\\
\vspace*{.5cm}
{\em School of Mathematics\\
Trinity College Dublin\\
Dublin 2, Ireland}\\
\vspace{2cm}

\begin{abstract}
We study the formation of BTZ black holes by the collision of point 
particles.  It is shown that the Gott time machine, originally constructed 
for the case of vanishing cosmological constant, provides a precise
mechanism for black hole formation.  As a result, one obtains
an exact analytic understanding of the Choptuik scaling. 
\end{abstract}
\vspace{1cm}
August 1999
\end{center}
\end{titlepage}

In \cite{Gott}, a precise mechanism was presented for the production
of closed timelike  curves (CTC's).  In particular, the spacetime of two
point particles with mass and boost parameters $\alpha$ and $\xi$,
in $(2+1)$-dimensional spacetime with vanishing cosmological
constant $\Lambda$, was shown to produce CTC's 
if the inequality, $\sin \frac{\alpha}{2}\;\cosh \xi > 1$, 
is satisfied. 
In \cite{Carr1, DJH1}, this Gott time machine was analysed
in terms of the group theoretic approach to point particles \cite{DJH2}.
One essential feature of this approach is that the point particle spacetime
is obtained via a suitable identification by an elliptic
(timelike) generator of the isometry group. 
It was demonstrated in \cite{Carr1,DJH1} 
that the effective two-particle generator 
(the Gott time machine) becomes hyperbolic (spacelike)
precisely when the Gott condition is satisfied. Various
arguments were then presented to ensure that such
CTC's do not arise in physically acceptable spacetimes 
\cite{Carr1, DJH1, Carr2, Menotti}.
The possibility that black hole formation could provide an escape route 
from the CTC's was alluded to in \cite{Gott}. However, the absence of
a black hole spacetime with $\Lambda = 0$ means that this route 
for chronology protection is unavailable.

It is noteworthy that in $(2+1)$-dimensional anti-de Sitter gravity
($\Lambda < 0$), we do indeed
have the Ba\~{n}ados-Teitelboim-Zanelli (BTZ) black hole 
solution \cite{BTZ1,BTZ2}, for a review
see \cite{Carlip1}. From our point of view, the most important aspect
of this black hole is that it is defined by a hyperbolic isometry.
One may choose a fundamental region
of this hyperbolic isometry, and define the black hole spacetime
by identification of the region's boundaries by the isometry.
Equivalently, the black hole spacetime may be defined as
the quotient of three-dimensional anti-de Sitter space
$({\mathrm{adS}}_{3})$ by the cyclic group generated by the
hyperbolic isometry \cite{BTZ2}.
In this paper, we show that the Gott time machine, suitably extended 
to the anti-de Sitter 
case, is precisely the mechanism for BTZ black hole formation 
in particle collisions.
If the Gott condition is satisfied, the effective two-particle generator 
becomes hyperbolic.
Thus, one can immediately identify the resulting
quotient spacetime as the BTZ black hole.
This understanding of black hole formation via the Gott time machine
is based simply on the observation that the formation of the black hole
is effected by the transition from two elliptic (particle)
generators to a hyperbolic (black hole) generator.

Having understood BTZ black hole formation at the algebraic level, 
it is of interest to see if this sheds light on the origin of Choptuik 
scaling \cite{Chop},
for a review see \cite{Gund}.
Interest in this critical phenomenon is centered around the
fact that the threshold
for black hole formation in the space of initial
data has a simple structure. In particular, the black hole
parameters exhibit universal power-law scaling behaviour.
We demonstrate that the natural order parameter for BTZ black hole 
formation is the trace of the associated generator. 
In this way, we gain an
exact analytic understanding of Choptuik scaling
in $2+1$ dimensions. 
It is then a simple matter
to read off the Choptuik scaling, which has an exponent $\gamma = 1/2$. 

We recall \cite{Carlip1} 
that $\mathrm{adS}_{3}$ can be viewed as the group manifold of
$SL(2,\mathbf{R})$, with isometry group  
$(SL(2,{\mathbf{R}}) \times SL(2,{\mathbf{R})})/Z_{2}$. 
Thus, for ${\mathbf{X}} \in SL(2,{\mathbf{R}})$, the isometry group
acts by left and right multiplication, ${\mathbf{X}} \rightarrow \rho_{L}
{\mathbf{X}}\rho_{R}$, with the identification  
$(\rho_{L},\rho_{R}) \sim (-\rho_{L}, -\rho_{R})$.
We shall use the equivalent $SU(1,1)$ notation instead of 
$SL(2,{\mathbf{R}})$;
they are related by conjugation, which is given explicitly in \cite{Steif}.  

As shown in \cite{DJH2}, the spacetime for a single static point particle 
with $\Lambda = 0$ is obtained by removing a wedge of deficit
angle $\alpha$, and identifying opposite sides
of the wedge. The  particle spacetime is defined via
the rotation generator with angle
$\alpha$,
\bea
R(\alpha) =
\left (
\begin{array}{cc}
e^{-i \alpha/2}&0\\
0& e^{i\alpha/2}
\end{array}
\right ). 
\eea
The mass $m$ of the particle is given by $\alpha = \pi m$, 
in units with $8G=1$, and the resulting spacetime has  a naked conical 
singularity.

A moving particle is obtained by boosting to the rest frame of the particle,
rotating, and then boosting back. The corresponding boost matrix is
\bea
B(\mbox{\boldmath $\xi$}) =
\left (
\begin{array}{cc}
\cosh \frac{\xi}{2}&e^{-i\phi} \sinh \frac{\xi}{2}\\
e^{i\phi} \sinh \frac{\xi}{2}& \cosh \frac{\xi}{2}
\end{array}
\right ), 
\eea
where $\mbox{\boldmath $\xi$}$ is the boost vector 
with $\xi = |\mbox{\boldmath $ \xi$}|$, and
$\phi$ is the polar angle.
Thus, the generator for a moving particle is 
\bea
T = B(\mbox{\boldmath $\xi$}) R(\alpha) B^{-1}(\mbox{\boldmath $\xi$}),
\eea
where
\bea
T_{11} &=& e^{-i\alpha/2}\left[ \cosh^{2}\frac{\xi}{2} - e^{i\alpha}\;
\sinh^{2}\frac{\xi}{2}\right],\non\\
T_{12} &=& 2i\;\sin\frac{\alpha}{2}\;e^{-i\phi}\;\cosh\frac{\xi}{2}\;
\sinh\frac{\xi}{2},\non\\
T_{21} &=& T_{12}^{*}, T_{22}= T_{11}^{*}. 
\label{T}
\eea
Since our considerations rest in anti-de Sitter space, we simply note
that the static and moving particle spacetimes are
defined in an analogous fashion, with left and right generators.
Particle spacetimes for non-zero cosmological constant have been
considered in \cite{DJ}.

To construct the Gott time machine, we consider a two-body collision
process, with particles labelled by $A$ and $B$. The effective two-particle
generator is then the product \cite{Carr1, DJH1, DJH2}, namely
$T^{G} = T_{B}T_{A}$.
The central object of interest to us is the trace of this generator. Using
(\ref{T}), it is straightforward to compute
\bea
\frac{1}{2}\; {\mathrm{Tr}}\;T^{G} &=& \cos \frac{\alpha_{A}}{2}\;\cos
\frac{\alpha_{B}}{2} + \sin \frac{\alpha_{A}}{2}\;\sin\frac{\alpha_{B}}{2}
\non\\
&-& \sin\frac{\alpha_{A}}{2}\;
\sin\frac{\alpha_{B}}{2}\left[ \cosh^{2}\left(\frac{\xi_{A} 
+ \xi_{B}}{2}\right) 
+\cosh^{2}\left(\frac{\xi_{A} - \xi_{B}}{2}\right)\right]\non\\
&+& \sin\frac{\alpha_{A}}{2}\;
\sin\frac{\alpha_{B}}{2}\cos(\phi_{A} - \phi_{B})
\left[ \cosh^{2}\left(\frac{\xi_{A} + \xi_{B}}{2}\right) 
-\cosh^{2}\left(\frac{\xi_{A} - \xi_{B}}{2}\right)\right].
\eea
The original Gott time machine is recovered by choosing particles
with equal masses, and equal and opposite boosts, namely
$\alpha_{A} = \alpha_{B} = \alpha, \xi_{A} = \xi_{B} = \xi, 
\phi_{A} - \phi_{B} = \pi$.
We find  
\bea
\frac{1}{2}\; {\mathrm{Tr}}\;T^{G} = 1 - 2 \sin^{2}\frac{\alpha}{2}\;
\cosh^{2}\xi.
\label{Gott}
\eea

The significance of this result can be  appreciated by
recalling that
the isometries of ${\mathrm{adS}}_{3}$ are classified according to 
the value of their trace. 
We have
\bea
\mid{\mathrm{Tr}}\; T\mid &<& 2, \;\; 
{\mathrm{Elliptic}},\non\\
\mid{\mathrm{Tr}}\; T\mid &=&2, \;\; 
{\mathrm{Parabolic}},\non\\
\mid {\mathrm{Tr}} \;T\mid &>& 2, \;\; {\mathrm{Hyperbolic}}.
\eea 
Thus, when the Gott condition is satisfied, 
we have $\sin^{2}\frac{\alpha}{2}\;\cosh^{2} \xi > 1$, and thus $T^{G}$
is a hyperbolic generator, and when
$\sin^{2} \frac{\alpha}{2} \;\cosh^{2} \xi < 1$, we have
an elliptic generator.

Armed with this observation, we are now in a position
to discuss the implications of the Gott time machine for
black hole formation.
First, we recall that the conventional mass parameter of the BTZ 
black hole is denoted by $M$, while the point particle mass $m$
is related by 
$m = 2 (1 - \sqrt{-M})$.
As a result, the point particle mass spectrum is $-1 < M < 0$,
while the black hole mass spectrum is $M \geq 0$, with $M=-1$
corresponding to ${\mathrm{adS}}_{3}$. 

Let us first consider the static black hole case, in which the left
and right generators are taken to be equal \cite{Carlip1}.
We recall also that the isometries of ${\mathrm{adS}}_{3}$
are subject to the identification $(\rho_{L}, \rho_{R}) \sim
(-\rho_{L}, -\rho_{R})$.
Thus, we may take $\rho_{L} = \rho_{R} = -T^{G} \equiv \rho$. 
The BTZ black hole is defined as the quotient of 
${\mathrm{adS}}_{3}$ by a cyclic group with a single hyperbolic 
generator \cite{BTZ2}. If the Gott condition is satisfied,
then as we have seen $T^{G}$ is a hyperbolic generator. Thus, 
the Gott time machine results in BTZ black hole formation.
The black hole mass is then given by \cite{Carlip1, Steif}
\bea
\frac{1}{2}\; {\mathrm{Tr}}\;\rho &=& \cosh \pi \sqrt{M} = -1 + 2 \sin^{2}
\frac{\alpha}{2}\;\cosh^{2}\xi \equiv p,
\label{mass}
\eea
where $p \geq 1$.

It is important to consider the defining equation
of this process, namely
\bea
T_{B}T_{A} = T^{G}.
\label{Gott2}
\eea
On the left-hand side, we have the input data given by the particle
mass and boost parameters $\alpha$ and $\xi$. The incoming
particles $A$ and $B$ have been set up in a symmetrical way, with
equal mass and boost parameters, and the timelike geodesics
representing their worldlines will
intersect at a given time, say $t=0$. Thus, we may regard
this as the time of collision of the two particles.
The product of
the particle generators represents the effective generator of the system
at this time \cite{Mats1}. As we have seen, the effective generator
at the time of collision is hyperbolic if the Gott condition is
satisfied. We may then interpret equation (\ref{Gott2})
as defining the formation of a BTZ black hole
at time $t=0$, with the value of the black hole mass
fixed by the input parameters
$\alpha$ and $\xi$. Thus, equation (\ref{Gott2}) encodes
dynamical information.
However, the precise details of the
motion of the particles corresponding to the generators $T_{A}$ and $T_{B}$
prior to the collision, as well as the motion after collision
may also be studied. Indeed, this
analysis has been performed for massless particles in \cite{Mats1,Mats2}.

We see from (\ref{mass}) that the natural order parameter
for black hole formation in $(2+1)$-dimensional anti-de Sitter 
gravity is
the trace of the generator. This takes a critical value
at the threshold for black hole formation, corresponding to the 
critical value of the parameter $p_{*} = 1$.
Clearly, $p=p_{*}$ corresponds to the black hole vacuum $M=0$,
where the Gott generator is parabolic. 
Since the parameter $p$ depends on the initial data $\alpha$ and $\xi$,
we can read off the critical boost $\xi$ for any given
mass $\alpha$. As an example, taking $\alpha = \pi/3$, we have black hole
formation when $\cosh\xi \geq 2$.
We have 
\bea
\pi \sqrt{M} = {\mathrm{arccosh}}\; p
= \ln\left[p + \sqrt{p^{2} - 1}\right].
\label{exact}
\eea
We stress that the above expression is an exact analytic formula
for the formation of a BTZ black hole
in terms of the input (initial) parameters $\alpha$ and $\xi$,
equivalently $p$.
From this, we can immediately determine the Choptuik scaling, by
studying the behaviour near $p_{*}$. 
The mass $M$ and horizon length $r_{+}$ are related by
$\sqrt{M} = r_{+}/l$, where $l$ is the scale of ${\mathrm{adS}}_{3}$.
Writing $p = p_{*} + \epsilon$, we find to leading order 
\bea
\frac{r_{+}}{l} = \frac{\sqrt{2}}{\pi} (p - p_{*})^{1/2}.
\eea
Thus, we observe a scaling factor of $ \gamma = 1/2$.
Note that this is indeed a universal scaling since BTZ formation
always requires a hyperbolic generator.
The universal scaling value of $1/2$ is simply a consequence
of the fact that the mass depends on the inverse cosh function.
We also note that the derivative of the order parameter 
diverges at the critical value $p_{*}$.

If the Gott condition is not satisfied, then one has an
effective  particle
spacetime with an elliptic generator, whose effective deficit angle is 
denoted by $\alpha_{\mathrm{eff}}$. This can be obtained by  
continuation of (\ref{mass}) to negative $M$ values. We find,
\bea
\frac{1}{2}\; {\mathrm{Tr}}\; \rho &=& \cos  \pi \sqrt{-M} = 
-1 + 2 \sin^{2}\frac{\alpha}{2}\;\cosh^{2}\xi \equiv p,
\eea
where now $p < p_{*}$. 
Once again, we have an exact analytic expression for the 
mass parameter on the other side of the transition, and of course
the Choptuik scaling exponent is again $\gamma = 1/2$, with
\bea
\alpha_{\mathrm{eff}} = 2 \pi - 2 \sqrt{2}(p_{*} - p)^{1/2}.
\eea
While the Choptuik scaling is evident near the black hole
threshold, we stress that in this model we have recourse
to the exact  analytic expression (\ref{exact}).

To obtain the spinning BTZ black hole, we simply need to 
have independent left and right generators, and these are also
given in terms of the Gott generators.
Taking 
$\alpha_{A} = \alpha_{B} = \alpha, \phi_{A} - \phi_{B} = 0$, 
we find the left generator $\rho_{L} = -T^{G}$, with
\bea
\frac{1}{2} \;{\mathrm{Tr}}\;\rho_{L} = \cosh\left
(\frac{\pi}{l}(r_{+} - r_{-}) \right) = -1 + 2 \sin^{2}\frac{\alpha}{2}\;
\cosh^{2}\left(\frac{\xi_{A} - \xi_{B}}{2}\right) \equiv \tilde{p}.
\eea
For the right generator, we choose
$\alpha_{A} = \alpha_{B} = \alpha, \phi_{A} - \phi_{B} = \pi$,
with $\rho_{R} = -T^{G}$, leading to
\bea
\frac{1}{2} \;{\mathrm{Tr}}\;\rho_{R} = \cosh \left( \frac{\pi}{l}
(r_{+} + r_{-})\right)
= -1 + 2 \sin^{2}\frac{\alpha}{2}\;
\cosh^{2}\left(\frac{\xi_{A} + \xi_{B}}{2}\right) \equiv p.
\eea
The input data for particles $A$ and $B$ is no longer
symmetric, but nevertheless $\rho_{L}$ and $\rho_{R}$ both
become hyperbolic if the input parameters $\alpha, \xi_{A}, \xi_{B}$
satisfy the appropriate Gott conditions.
In the above, we have introduced the parameters $r_{\pm}$ which denote
the locations of the inner and outer horizons \cite{Carlip1}.
They are related to the mass $M$ and angular momentum $J$ 
by $M = (r_{+}^{2} + r_{-}^{2})/l^{2}, J = 2r_{+}r_{-}/l$.
In this case, we obtain Choptuik scaling for both 
mass and angular momentum in the form 
\bea
\frac{r_{+} - r_{-}}{l} &=& \frac{\sqrt{2}}{\pi} (\tilde{p} - 
\tilde{p}_{*})^{1/2},\non\\
\frac{r_{+} + r_{-}}{l} &=& \frac{\sqrt{2}}{\pi} 
(p - p_{*})^{1/2}.
\eea

As pointed out in \cite{BKSW}, the Euclidean BTZ black hole is a 
geometrically finite Kleinian manifold with the topology of a solid
torus. In essence, this allows one to invoke a theorem of 
Sullivan \cite{Sull, McM} 
to establish a no hair theorem for the Euclidean black hole.
As a result,  the black hole and its hyperbolic generators are described
by at most two parameters.
Assuming that the continuation
to Lorentzian signature does not induce additional parameters,
one concludes that the BTZ black hole is described by mass and angular
momentum only. 
One can consider many-body extensions of the Gott time machine.  
By the above reasoning, these will
lead to the formation of a spinning BTZ black hole, if there are Gott
conditions
which produce independent left and right hyperbolic generators.

In conclusion, we have shown that the original Gott time machine
provides a precise mechanism for BTZ black hole formation in
particle collisions.
The problems of CTC's and chronology protection \cite{Haw}
are overcome by the creation of the black hole horizon, as 
soon as the Gott condition is satisfied. This is indeed a satisfying scenario.
We mention that the recent study of BTZ
black hole formation from massless particle collisions \cite{Mats1,Mats2}
is based on the lightlike
analogue of the Gott time machine \cite{DS}, and thus is also
guaranteed to produce the hyperbolic
generator necessary for BTZ formation. 
The holographic description of this creation process has been 
investigated within the ${\mathrm{adS}}/{\mathrm{CFT}}$
correspondence in \cite{Bal}. 
A Gott time machine in anti-de Sitter space was also studied in 
\cite{Holst}.
By definition, the BTZ black hole is defined as the
quotient spacetime of a hyperbolic generator.
Thus, irrespective of which type of matter is used to produce such a 
black hole,
the ultimate result is that the mass of the black hole is defined
via the trace of the hyperbolic generator. Hence, Choptuik scaling
with a critical exponent
$\gamma = 1/2$ will always be present. Indeed, this
exponent was observed for collapsing dust shells in \cite{Peleg}.

\noindent {\large \bf Acknowledgements}\\
This work is part of a project
supported by Enterprise Ireland Basic Research Grant SC/98/741.
D.B. would like to thank the Theory Division at CERN for
hospitality, and Luis Alvarez-Gaum\'{e} for valuable discussions.

\end{document}